# Nanophotonics with 2D Transition Metal Dichalcogenides

**ALEX KRASNOK,**[1] **SERGEY LEPESHOV,**[2] **AND ANDREA ALÚ**[1]

[1] *Department of Electrical and Computer Engineering, The University of Texas at Austin, TX 78712, USA*
[2] *ITMO University, St. Petersburg 197101, Russia*

**Abstract:** Two-dimensional transition metal dichalcogenides (TMDCs) have recently become attractive semiconductor materials for several optoelectronic applications, such as photodetection, light harvesting, phototransistors, light-emitting diodes and lasers. They are particularly appealing because their bandgap lies in the visible and near-IR range, and they possess strong excitonic resonances, high oscillator strengths, and valley-selective response. Coupling these materials to optical nanocavities enhances the quantum yield of exciton emission, enabling advanced quantum optics and nanophotonic devices. Here, we review state-of-the-art advances on hybrid exciton-polariton structures based on monolayer TMDCs coupled to plasmonic and dielectric nanocavities. We first generally discuss the optical properties of 2D $WS_2$, $WSe_2$, $MoS_2$ and $MoSe_2$ materials, paying special attention to their energy and photoluminescence/absorption spectra, excitonic fine structure, and to the dynamics of exciton formation and valley depolarization. We then discuss light-matter interactions in hybrid exciton-polariton structures. Finally, we focus on weak and strong coupling regimes in monolayer TMDCs-based exciton-polariton systems, envisioning research directions and future opportunities based on this novel material platform.

## References and links

## 1. Introduction

In recent years, researchers all over the world have made important steps towards the development of new materials with unique optical and electronic properties. One prominent example consists of atomically thin single layered (1L) transition metal dichalcogenides (TMDCs) [1–9]. Monolayer TMDCs are formed by a hexagonal network of transition metal atoms (Mo, W) hosted between two hexagonal lattices of chalcogenide atoms (S, Se). The resulting materials have the common formula $MX_2$ (where M is Mo or W and X is S or Se).

Electronically, 1L-TMDCs behave as two-dimensional semiconductors, with bandgaps lying in the visible and near-IR range. This property profitably distinguishes them from other 2D materials, like graphene and hBN, whose bandgaps occur at longer wavelengths. In the monolayer limit, the bandgaps of these materials are direct, enabling enhanced interactions of dipole transitions with light. Reduced dielectric screening and strong Coulomb interactions between charged particles (electrons and holes) result in strong excitonic resonances in the visible and near-IR range with large binding energies. Strong Coulomb interactions in these materials lead to the formation of strongly bound excitons (binding energies of 0.2 to 0.8 eV) [10–14], charged excitons (trions) [15–17], and excitonic molecules (biexcitons) [19–23]. Localized and dark excitonic states in 1L-TMDCs have also been demonstrated [19,24,25]. Inversion symmetry breaking and strong spin-orbital coupling in 1L-TMDCs lead to direct bandgap transitions occurring at energy-degenerate K (K') points at the edges of the 2D hexagonal Brillouin zone, enabling valley-selective circular dichroism [25–30]. This effect has become of great importance in the emerging field of *valleytronics*, which exploits valley pseudo-spin manipulation to transmit information [7].

Due to the monolayer nature of 1L-TMDCs, their high oscillator strength and the potential for tuning, these materials have become a unique class of 2D materials for optoelectronic applications, such as photodetection and light harvesting [31–34], phototransistors and modulation [35,36], light-emitting diodes [37–39] and lasers [40,41]. For example, the electroluminescence from light-emitting diodes (LEDs) based on $MoS_2$ [42,43], $WSe_2$ [44], and $WS_2$ [45] has been recently demonstrated. Operation of 1L-TMDCs in single photon emission regime has also been reported [46]. The quantum yield of emission (see Section 2) of these materials depend on many factors, including fabrication techniques (mechanical exfoliation or chemical vapor deposition), type of substrate, defects and so on. Typically, as-prepared 1L-TMDCs have a relatively low quantum yield of ~0.1-10%. These low values can be significantly improved (up to ~95%) by chemical treatment by organic superacids [47,48].

Another way to improve 1L-TMDC emission properties consists in coupling them with specifically tailored single resonant optical nanocavities and arrays of them, in the form of metasurfaces [24,49–63]. A variety of new optical effects stemming from the interaction of 1L-TMDCs with plasmonic (i.e., made of noble metals) and high-index dielectric (Si, Ge or GaP) nanocavities has been demonstrated. Examples include the observation of *strong plasmon-exciton coupling* [50,57,64,65], pronounced *Fano resonances* [59] and plasmon-induced *resonance energy transfer* [66], which are very attractive for various quantum optics and nanophotonic applications. These effects benefit from squeezed states of light within plasmonic and dielectric resonators, enabled by their small mode volumes and the strong dipole moment of excitons in 1L-TMDCs. Resonant nanophotonic structures can also alter 1L-TMDCs emission properties, for example brighten the so-called dark excitons [24,51], which poorly radiate without this essential coupling with optical resonators. In addition, nanophotonic structures may improve other inherently weak properties of 1L-TMDCs, such as exciton energy transfer, which is typically limited to a short range of ~1 μm. Recently, it has been demonstrated that the exciton energy transfer range can be extended to tens of microns in hybrid structures mediated by an exciton-surface plasmon polariton-exciton conversion mechanism [67].

In the following, we provide an overview of the state-of-the-art advances of hybrid exciton-polariton structures based on 1L-TMDCs coupled to plasmonic and dielectric nanocavities. In Section 2, we discuss the optical properties of 2D $WS_2$, $WSe_2$, $MoS_2$ and $MoSe_2$ materials, with special attention to their energy spectra, photoluminescence and absorption spectra, excitonic fine structure, and dynamics of exciton formation and valley depolarization. Then in Section 3, we describe suitable optical resonances of plasmonic and dielectric nanoantennas that may be employed to enhance TMDCs. In Section 4, we provide a theoretical background to describe light-matter interactions in hybrid exciton-polariton structures, focusing on weak and strong coupling regimes of light-meter interaction. Finally, in Section 5, we discuss these regimes in 1L-TMDCs based exciton-polariton systems, emphasizing various practical realizations.

## 2. Optical properties of 1L-TMDCs

Single layered Mo and W-based TMDCs ($MoS_2$, $MoSe_2$, $WS_2$, $WSe_2$) transition metal dichalcogenides behave as two-dimensional semiconductors with their bandgaps lying in visible and near-IR ranges [2]. This property distinguishes 1L-TMDCs from other known 2D materials like graphene (semimetal, no bandgap) and hBN (insulator, the bandgap is ~6 eV). In the monolayer limit, TMDCs are particularly interesting because their bandgaps become direct, thus enabling enhanced interactions of dipole transitions with light [68]. The transition to a direct bandgap semiconductor makes it easier for photons with an energy equal to the bandgap to be absorbed or emitted, whereas indirect bandgap materials require additional phonon absorption or emission to compensate for the difference in momentum, making absorption less efficient. Additionally, the reduced dielectric screening and strong Coulomb interactions result in formation of strongly bounded excitons [10–14], charged excitons (trions) [15–17], excitonic molecules (biexcitons) [18–22], localized excitons [18] and dark excitons [24] with large binding energies. For example, the binding energies of $X_A$ excitons in 1L-MoS$_2$ [10], MoSe$_2$ [14], WSe$_2$ and WS$_2$ [12,13,69] were measured to be in the range of 0.2 to 0.8 eV. The absorption coefficients of 1L-TMDCs can reach up to $1.0 \times 10^6$ cm$^{-1}$, which is much higher than those of conventional quantum dots ($0.2 \times 10^5$ cm$^{-1}$ for CdSe) [70], and comparable to dense J-aggregates ($0.5 \times 10^6$ cm$^{-1}$) [71]. The typical thickness of 1L-TMDCs is ~0.7 nm [2,8] is only ~3 times larger than the thickness of graphene.

Direct band gap transitions in 1L-TMDCs occur at the energy-degenerate K (K') points, at the edges of the 2D hexagonal Brillouin zone. Due to inversion symmetry breaking and strong spin-orbital coupling in 1L-TMDCs, the electronic states of the two valleys have different chirality, which leads to valley-selective circular dichroism [25–30]. This effect is key for valleytronics applications, which focus on the manipulation of valley pseudo-spins to encode signals and information [7]. Despite a tremendous interest in this area, valleytronics faces the obstacle of short valley depolarization times, only ~5 ps for 1L-TMDCs at cryogenic temperatures and sub-picosecond at room temperatures. Inversion symmetry breaking and strong spin-orbital coupling also lead to so-called spin-forbidden dark excitons polarized along the out-of-plane direction, which can be brightened by strong static magnetic fields [72] or tip-enhanced Purcell effect [24].

Below we review the optical properties of 1L-TMDCs with focus on WS$_2$ and WSe$_2$, which are very often used for LED and laser technologies. Almost any conclusion drawn in the following is also applicable to other 1L-TMDC materials, because of their similar nature. We must stress that the optical properties of 1L-TMDCs (PL spectra, Raman spectra, transmission/absorption spectra) may slightly vary as a function of several factors, including fabrication techniques (mechanical exfoliation or chemical vapor deposition), type of substrate, defects and others.

*2.1 Optical properties of 2D WS$_2$, photoluminescence tuning and excitonic fine structure*

1L-tungsten disulfide (WS$_2$) is one of the most studied 2D transition-metal dichalcogenide semiconductors [19,22,64,73–77]. 1L-WS$_2$ has been attracting growing interest due to its unique properties, such as large spin-orbit coupling, high emission quantum yield [78], large exciton/trion binding energies [13,79], and demonstrated nonblinking photon emission [69]. Additionally, this material is attractive because it enables strong-coupling with excitons [57,64,75] and trions [17], as discussed in more detail in Section 5 in the context of tuning PL spectra and excitonic fine structure (trions, biexcitons). 1L-WS$_2$ contains a single layer of W atoms with 6-fold coordination symmetry, hexagonally packed between two trigonal atomic layers of S atoms. The energy band diagram (energy versus wavevector k) of 1L-WS$_2$, as well as its absorption spectra (orange curve) and photoluminescence (blue curve), are shown in Figs. 1(a), (b). The direct bandgap of 1L-WS$_2$ is ~2.1 eV [45,81]. Inversion symmetry breaking of the crystal lattice combined with strong spin-orbit coupling results in a large valence band splitting (~427 meV) at the K (K') points in the first Brillouin zone, Fig. 1(a). This phenomenon gives rise to two different valley excitons ($X_A$ and $X_B$ excitons), which are associated with optical transitions from the upper and the lower valence band to the bottom of the conduction band, respectively. The $X_A$ exciton binding energy was demonstrated to be equal ~0.71 eV [79]. Note that the exciton binding energy may strongly depend on TMDC fabrication technique, as well as the defect and exciton concentrations [82]. It causes an appearance of two resonances in the absorption spectrum, Fig. 1(b). The PL spectrum also consists of two resonances corresponding to $X_A$ and $X_B$ excitons, but the PL emission from $X_B$ exciton is very weak [right inset in Fig. 1(b)].

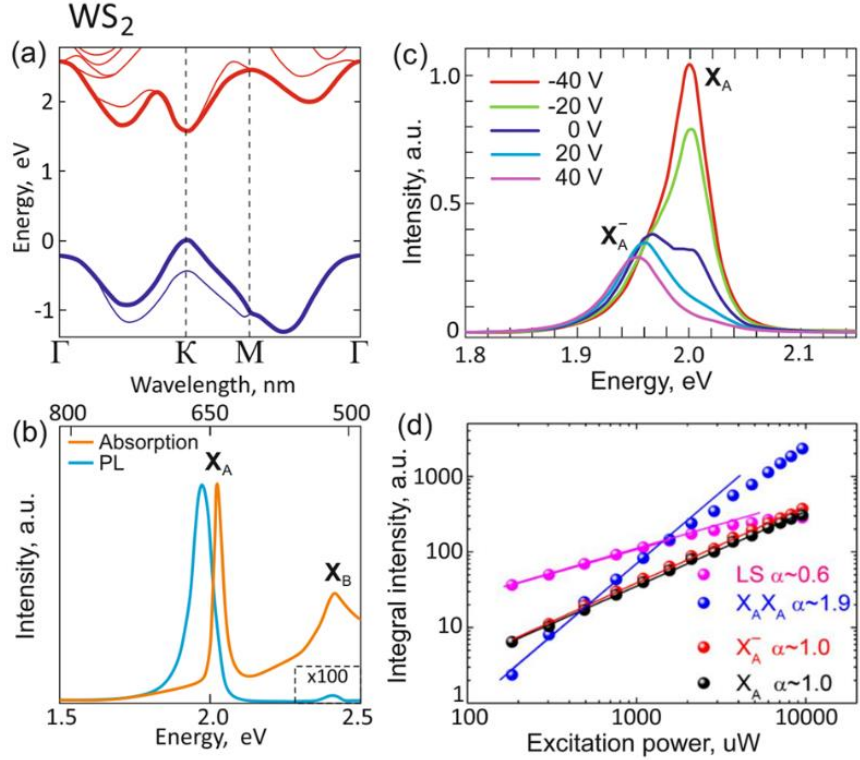

Fig. 1. Optical properties of 1L-WS$_2$. (a) Energy structure of 1L-WS$_2$, blue and red curves demonstrate the valence and conduction bands. (b) Absorption spectra (orange curve) and photoluminescence (blue curve) of 1L-WS$_2$ at room temperature. (c) Photoluminescence spectra at the back-gate voltages between -40 and +40 V at room temperature. (d) Log-log plots of integrated intensities of four emission components versus excitation power at 4.2K. Solid lines are linear fit curves, and the corresponding $\alpha$ values in the legend are the slopes. Adapted with permission from: (a),(b) – Ref. [80], (c),(d) – Ref. [18].

It has been demonstrated that WS$_2$ PL spectrum can be tuned in various ways, including chemical doping [74], mechanical strain [83], type of substrate [84], surrounding dielectric environment [85], electrical doping [18,73], laser intensity [18,82,86], and interaction with optical cavities [17,87,88]. For example, in [18] tunable emission in 1L-WS$_2$ was demonstrated by controlling the electrical doping strength. Fig. 1(c) shows the PL spectra of 1L-WS$_2$ at room temperature for different gate voltages, ranging from -40 to 40 V, continuously modulating the charge carrier density. These results demonstrate that the exciton peak has two components, the neutral exciton $X_A$ and the charged exciton $X_A^-$ (trion), located at lower energy. At cryogenic temperatures, trions can contribute significantly to the PL emission spectra [17]. Note that the doping by free charge carriers can occur because of WS$_2$ interaction with the dielectric environment (for example, the substrate) [89].

At low temperatures, other electron-hole states can be observed. Fig. 1(d) presents the integrated PL emission intensities of $X_A$ (excitons), $X_A^-$ (negatively charged trions), $X_A X_A$ (biexcitons), and LS (localized excitons) from 1L-WS$_2$ as a function of the excitation power at 4.2K [18]. At low excitation intensities, LS, $X_A$, and $X_A^-$ give the main contribution to PL emission. The emission from exciton and trion states increases linearly ($\alpha \sim 1.0$) with the excitation power, whereas the emission from LS shows a sublinear dependence ($\alpha \sim 0.6$). On the contrary, the integrated intensity of biexcitons $X_A X_A$ grows quadratically with the excitation power ($\alpha \sim 1.9$), as expected. Similar results for biexciton emission from edges of triangular 1L-WS$_2$ have been demonstrated in [22]. Thus, biexciton PL emission dominates for high-power excitation, and it has been observed only at cryogenic temperatures (~4K). We note that the trion state can also have a finite structure with triplet and singlet trions, which can be observed in PL spectra at a temperature of 4K [90].

*2.2 Optical properties of 2D WSe$_2$. Dynamics of exciton formation and valley depolarization*

The second 1L-TMDCs material that we discuss is WSe$_2$, which has also been very well studied [11,91–94]. As applications of this material, in the following we will describe dynamics of exciton formation and valley depolarization. This material is also attractive for enabling strong-coupling regimes of light-matter interaction [50]. Also in this case, the conclusions about this material can be applied to other 1L-TMDC materials, except specific aspects of exciton-phonon interactions [76].

Figs. 2(a), (b) show the energy structure of 1L-WSe$_2$ [Fig. 2(a)] and the corresponding absorption spectra (orange curve) and photoluminescence (blue curve) [Fig. 2(b)]. The direct bandgap is ~1.6 eV. As in the case of 1L-WS$_2$, this material supports X$_A$ and X$_B$ excitons, which are associated with optical transitions from the upper and the lower valence band to the bottom of the conduction band. Accordingly, the absorption spectrum has two resonances corresponding to X$_A$ and X$_B$ excitons. Charged excitons in 1L-WSe$_2$ have also been observed [95]. It has been demonstrated that the PL spectra of 1L-WSe$_2$ can be tuned by the surrounding dielectric environment [96], electrical doping [97], magnetic field [98], laser intensity [91], and interaction with optical cavities [50].

For the full description of the temporal behavior of 1L-TMDCs like WSe$_2$, it is important to know their exciton and coherence lifetimes [99]. The exciton lifetime describes the average time during which an exciton exists, and it usually defines the excitonic spectral linewidth. The coherence time defines the time during which an exciton remembers the state of the excitation field (for example, polarization). The coherence time is usually less than the exciton lifetime and it is key for valleytronics and quantum optics applications [100].

In Ref. [91], a femtosecond optical-pump/mid-IR-probe (with probe energy ∼170 meV, which coincides with the difference between exciton eigenstates) was used to directly monitor the dynamics of photoexcited electron−hole pairs in 1L-WSe$_2$, Fig. 2(c). After highly nonresonant interband excitation (3.04 eV) by the femtosecond laser pulse, the concentration of free carriers increases, reaching its maximum after 0.5 fs. This excitation is followed by a rapid carrier relaxation towards the respective band minima. More than half of the carriers are bound into excitons already 0.4 ps after the excitation (Fig. 2(c), grey curve). The ratio between excitons and unbound electron−hole pairs increases up to 0.5 ps (Fig. 2(c), blue curve). Interestingly, the exciton concentration grows even after 0.5-0.6 ps, when the free carrier concentration starts to decay, and continues up to 1 ps. Then, both concentrations decay on a time scale of a few picoseconds, while a significant fraction of free carriers is still observed after 5 ps. These experiments have been performed at room temperature and ambient conditions.

These relatively short characteristic times $\tau^{loss}$ of exciton dynamics in 1L-TMDCs (1-5 ps) are caused by nonradiative processes, like exciton dissociation by defects and phonons, as well as possible four-body interactions [101], whereas the radiative exciton recombination is characterized by much longer times ($\tau^{rad}$ ~ 0.1−1 ns) [102]. The mismatch between these times causes a low quantum yield in emission, which can be defined as $\eta_0 = \gamma_{ex}^{0,rad}/(\gamma_{ex}^{0,rad} + \gamma_{ex}^{0,loss})$, where $\gamma_{ex}^{0,rad} = 1/\tau^{rad}$ and $\gamma_{ex}^{0,loss} = 1/\tau^{loss}$ are radiative and nonradiative decay rates, respectively. For 1L-TMDCs the values of quantum yield have been estimated to be ~0.1-10%, which can be increased by chemical treatment [46,47] and strong Purcell effect via coupling to specially designed optical nanocavities (Section 4) [48–61].

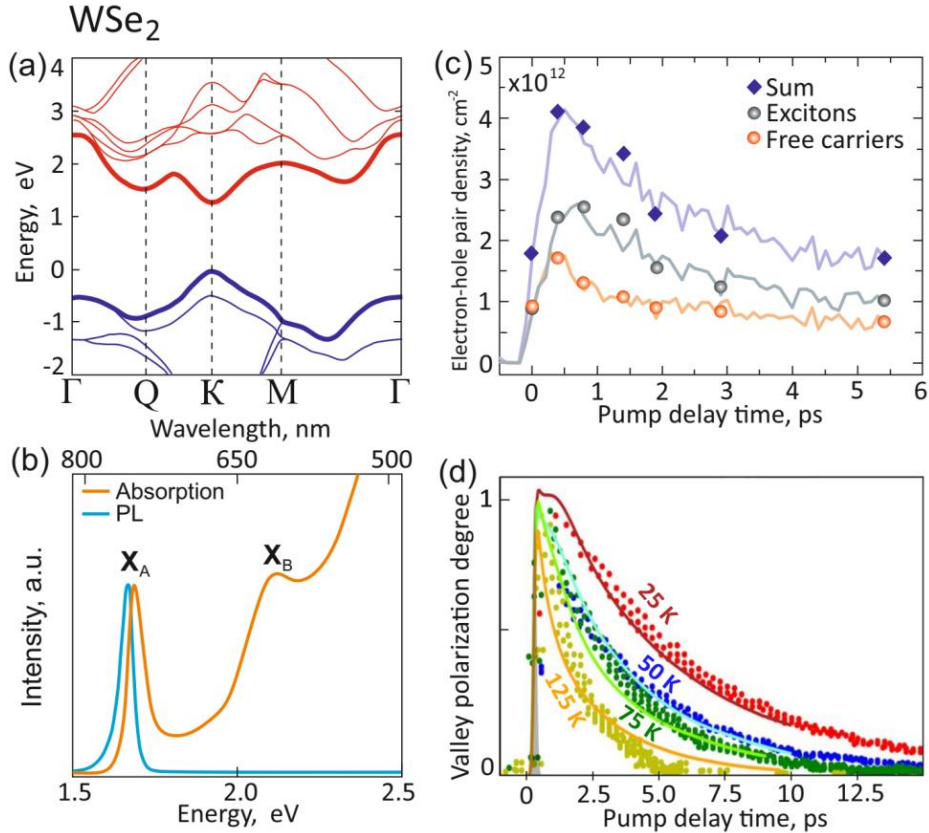

Fig. 2. Optical properties of 1L-WSe$_2$. (a) Energy structure; blue and red curves demonstrate the valence and conduction bands. (b) Spectra of absorption (orange curve) and photoluminescence (blue curve) of 1L-WSe$_2$ at room temperature. (c) Absolute electron-hole pair densities for 1s excitons (black spheres), unbound electron−hole pairs (red spheres), and total density of the two contributions (blue diamonds) as a function of the delay time after nonresonant excitation (3.04 eV). (d) The valley polarization degree as a function of delay time for several temperatures. The theory lines are compared with experiments from Ref. [103] (dots). The shadow gray region represents the pump pulse. Adapted with permission from: (a) – Ref. [91], (b) – Ref. [68], (c) – Ref. [91], (d) – Ref. [104].

The exciton valley dynamics of 1L-WSe$_2$ by pump-probe Kerr rotation technique has been experimentally studied in Ref. [103]. It has been demonstrated that the exciton valley depolarization time decreases significantly from ~6 ps up to ~1.5 ps when the lattice temperature increases from 4 K up to 125 K. In Ref. [104] these results have been obtained from *ab initio* calculations, which took into account all possible mechanisms of valley depolarization. The results are summarized in Fig. 2(d), where the valley polarization degree is shown as a function of the delay time for several temperatures. The results of theoretical calculations (lines) are compared with experimental results (dots) from Ref. [104]. The Kerr dynamics and its temperature dependence can be explained in terms of electron−phonon mediated processes that induce spin-flip intervalley transitions. Similar results have been demonstrated in Ref. [23].

### 2.3 Optical properties of 2D MoS$_2$ and MoSe$_2$

For the sake of completeness, we also summarize the fundamental optical properties of 1L-MoS$_2$ and MoSe$_2$ materials. Figs. 3(a),(c) show energy structure of 1L-MoS$_2$ and 1L-MoSe$_2$, respectively. Figs. 3(b),(d) show the absorption spectra (orange curve) and photoluminescence spectra (blue curve) of 1L-MoS$_2$ and 1L-MoSe$_2$, respectively. These materials also have $X_A$ and $X_B$ exciton resonances in their absorption spectra.

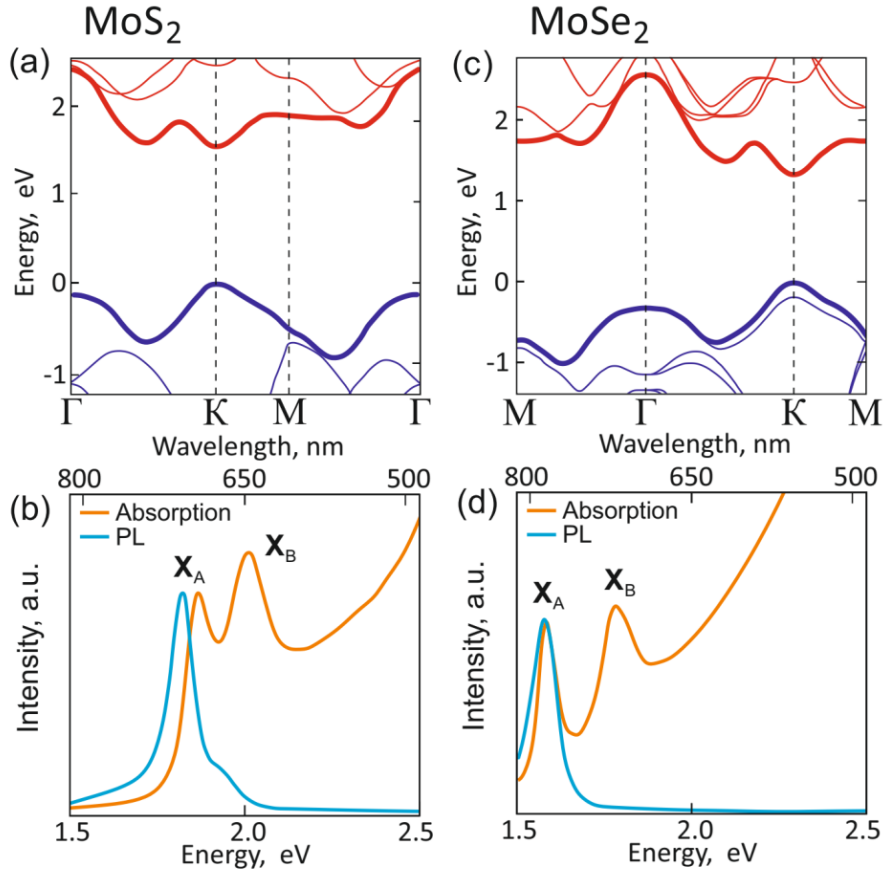

Fig. 3. Optical properties of 1L-MoS$_2$ and 1L-MoSe$_2$ (a), (c) Energy structure of 1L-MoS$_2$ and 1L-MoSe$_2$; blue and red curves demonstrate the valence and conduction bands. (b), (d) Spectra of absorption (orange curve) and photoluminescence (blue curve) of 1L-MoS$_2$ and 1L-MoSe$_2$. Adapted with permission from: (a) – Ref. [105], (b) – Ref. [53], (c) – Ref. [106], (d) – Ref. [52].

1L MoS$_2$ is a direct-gap semiconductor with a bandgap of ~1.8 eV [105,107]. The band-edge (X$_A$) exciton dominates the PL emission spectrum around 1.88 eV, Fig. 3(b). The binding energies of excitons and trions in 1L-MoS$_2$ were extracted to be around 900 and 40 meV, respectively. Notably, the trion/exciton PL peaks in 1L-MoS$_2$ materials can be tuned by different dielectric environments, with blue shifts of up to 40 meV and significant enhancement in PL emission intensities [108]. Relatively high exciton decay times of ~100 ps can be significantly reduced (up to ~10ps) by strong excitation [101].

In contrast, 1L MoSe$_2$ is a direct-gap semiconductor with a bandgap of ~1.65 eV [109]. Fig. 3(d) shows that the band-edge (X$_A$) exciton in the PL emission spectrum appearing at ~1.6 eV. At cryogenic temperatures, two features at ~1.65 and ~1.62 eV in this spectrum are attributed to the exciton and negative trion [21,110]. Trions and biexcitons in 1L MoSe$_2$ (at cryogenic temperatures) with large binding energies of 30meV (trions) and 50−70 meV (biexcitons) can be also resolved [16,111]. In addition to various LED and laser applications [37,42,112,113], 1L-MoS$_2$ and 1L-MoSe$_2$ have been used for thermal light emission [114], ultrasensitive photodetectors [31], exciton Hall effect [30], in electrically tunable exciton-plasmon polariton systems [115] and as a constitutive element in van der Waals heterostructures [116,117].

3. **Optical resonances of plasmonic and dielectric nanocavities**

Plasmonic nanocavities (NCs) made of noble metals (e.g., Au and Ag) provide efficient means to manipulate light and enhance light-matter interactions at the nanoscale [118–125]. Especially designed plasmonic nanocavities, such as single nanoparticles and arrays of them, can squeeze light in subwavelength dimensions, resulting in giant enhancement of interaction with quantum emitters, including 1L-TMDCs, quantum dots, defect centers in crystals, and molecules. Many interesting phenomena can be enabled by such interactions, including luminescence enhancement, ultrafast emission in the picosecond range, strong coupling, and surface-enhanced Raman scattering (SERS), all observed and applied in recent years to various real-life applications.

Dielectric nanocavities, made of high refractive index dielectrics such as Si, Ge or GaP, have also become interest for similar functionalities. Their resonant response stems from polarization charges, and thus it is less impacted by Ohmic losses compared to plasmonic phenomena [126–133]. In addition to low dissipative losses, dielectric NCs and nanostructures can provide resonant electric and magnetic optical responses, moderate light localization at the nanoscale, and strong Raman scattering, which are not typical for their plasmonic counterparts [132,133]. Because of their distinct properties, dielectric NCs have been proposed for high-harmonic generation [134–136], photonic topological insulators [137], and boosting the luminescence from various quantum emitters. Fano resonances [138], Purcell effect [139,140], and strong coupling [141] have been also demonstrated in dielectric NCs.

The optical response of spherical metallic and dielectric nanoparticles of radius $R$, as a basic example of efficient nanocavity, can be calculated based on Mie theory [142]. Normalized scattering $S_{sct}$, extinction $S_{ext}$ and absorption $S_{abs}$ cross-sections for nonmagnetic particles with dielectric permittivity $\varepsilon = n^2$ can be written as a sum of partial spherical waves with electric $a_l$ and magnetic $b_l$ scattering amplitudes

$$S_{sct} = \frac{2}{(kR)^2} \sum_{l=1}^{\infty} (2l+1)\left(|a_l|^2 + |b_l|^2\right),$$

$$S_{ext} = \frac{2}{(kR)^2} \sum_{l=1}^{\infty} (2l+1)\operatorname{Re}\left(a_l + b_l\right), \qquad (1)$$

$$S_{abs} = S_{ext} - S_{sct},$$

where $l$ defines the order of partial wave, $k$ is the wavenumber ($k = 2\pi n_h / \lambda$), $\lambda$ is the wavelength in vacuum and $\varepsilon_h = n_h^2$ is the dielectric permittivity of the surrounding medium. The electric and magnetic scattering amplitudes can be written as $a_l = \frac{R_l^{(a)}}{R_l^{(a)} + iT_l^{(a)}}$, $b_l = \frac{R_l^{(b)}}{R_l^{(b)} + iT_l^{(b)}}$, where $R_l$ and $T_l$ are expressed through Bessel $J_{l+1/2}(x)$ and Neumann $N_{l+1/2}(x)$ functions [142,143]. The equations $R_l^{(a)} + iT_l^{(a)} = 0$ and $R_l^{(b)} + iT_l^{(b)} = 0$ define the complex resonant frequencies $\omega_l$ of the electric and magnetic eigenmodes of the particle, respectively. Different resonant modes enumerated by $l$ are possible: electric (ED) and magnetic (MD) dipoles for $l = 1$ [corresponding to solutions of the a and b dispersion relation], electric (EQ) and magnetic (MQ) quadrupoles with $l = 2$, and so on. These eigenmodes manifest themselves at real frequencies as resonant enhancement of scattering and extinction cross-section spectra. In the general case, these expressions are satisfied at complex frequencies $\omega_l = \operatorname{Re}[\omega_l] + i\operatorname{Im}[\omega_l]$. Because the energy leaks out of the particle and is dissipated into absorption, the modes possess a finite lifetime $\gamma_{cav} = \operatorname{Im}[\omega_l]^{-1}$, usually known as *cavity decay rate*, which determines the half-width of resonance line at half maximum [144]. In general, the cavity decay rate includes both radiative and absorptive parts, $\gamma_{cav} = \gamma_{cav}^{rad} + \gamma_{cav}^{loss}$. The resonance is usually described by its *quality factor*, which is defined as $Q = \omega / \gamma_{cav}$. A relevant figure of merit for resonant NCs is their scattering efficiency, which can be defined as $\eta_{sct} = S_{sct}/S_{ext}$, and measures the portion of decay associated with radiation and scattering, which is desirable, over the one associated with absorption.

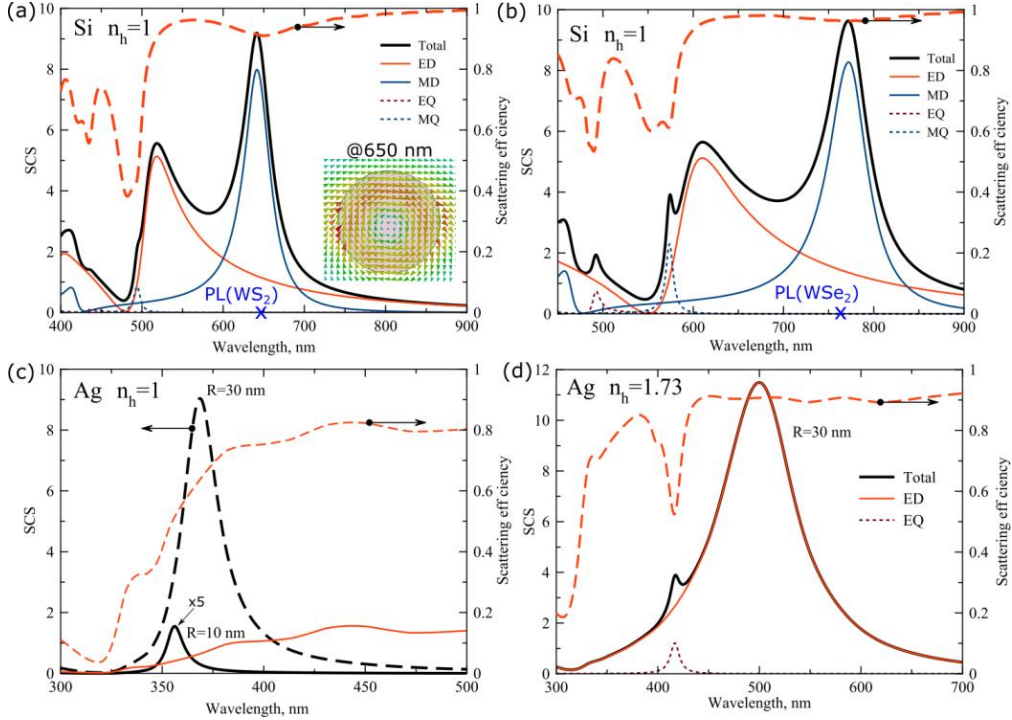

Fig. 4. Optical properties of dielectric (c-Si) and metallic (Ag) spherical nanoparticles. (a) Normalized total scattering cross-section (black curve) with partial contributions of ED (red solid curve), MD (blue solid curve), EQ (brawn dashed curve) and MQ (blue dashed curve) of a Si nanoparticle of R=80 nm in air $\varepsilon_h = 1$. Red dashed curve shows scattering efficiency. Inset: electric field distribution at MD resonance. (b) The same but for R=100 nm. (c) Normalized total scattering cross-section of the Ag nanoparticle of R=10 nm (black solid curve) and R=30 nm (black dashed curve) in air. Red curves show scattering efficiency. (d) Normalized total scattering cross-section of the Ag nanoparticle of R=30 nm (black solid curve) with partial contributions of ED (red solid curve) and EQ (brawn dashed curve) hosted in a dielectric media with $\varepsilon_h = 1.73$. Red dashed curve shows scattering efficiency.

Fig. 4 shows as an example the optical features of typically sized plasmonic and dielectric NCs. The resonant behavior of a dielectric NC strongly depends on its size, which may be used to tune its resonance to the response of an excitonic subsystem. For example, Figs. 4(a) and (b) demonstrate the optical properties of two spherical dielectric NCs (c-Si) with radius of 80 nm and 100 nm, respectively. These NCs have an MD resonance (blue solid curve) tuned to the PL maximum of, respectively, 1L-WS$_2$ and WSe$_2$. The scattering efficiency ($\eta_{sct}$) reaches 90% for an 80 nm-particle and increases up to 95% for the 100 nm-particle.

On the contrary, the plasmonic resonance (ED) of a plasmonic (Ag) NC weakly depends on the particle size, as it is mostly affected by the interface between metal and the surrounding material, and its curvature. Fig. 4(c) shows that an increase in particle radius from 10 nm to 30 nm redshifts the resonance only by ~10 nm, with associated decrease in its Q-factor from ~30 to ~10. The size increase turns the particle from mainly-absorptive (scattering efficiency ~5%) to mainly-scattering (scattering efficiency ~60%). The plasmonic resonance depends strongly on the permittivity of the dielectric environment, Fig. 4(d). For instance, an increase of $\varepsilon_h$ from 1 to 4 leads to a large redshift in the resonance peak from 370 nm (air) to 500 nm ($\varepsilon_h = 4$), with an increase of efficiency (up to 90%) for the R=30 nm particle, and a corresponding reduction in Q-factor to ~5.

The shape and arrangements of NCs are other convenient ways to tune their resonant properties. It has been demonstrated that the plasmonic resonance frequency can be engineered on demand by controlling these design parameters [145–147]. This mechanism for tunability is particularly important for 1L-TMDCs, since their absorption and PL peaks are usually significantly divided. In this case, the shape tunability of plasmonic NCs allows to adjust two (or even more) resonances to the absorption peak (for high excitation rate) and to the PL peak (high emission rate), simultaneously [53]. In the case of dielectric NCs, the shape tunability paves a way to control the total emission intensity through destructive interference in farfield of the radiation of two (or more) neighboring modes of different nature, with the realization of an almost nonradiative field configuration [148–150], leading to a large increase of Q factor and enhanced light-matter interactions.

### 4. Weak and strong coupling regimes

Coupling of an excitonic subsystem, such as 1L-TMDCs, with a NC can be described in the Jaynes-Cummings formalism [151] by the following matrix element of the coupling Hamiltonian: $H_{ij} = \mathbf{d} \cdot \mathbf{E}_v(\mathbf{r}_0) = \hbar g$, with $\mathbf{d} = e\langle j|\hat{\mathbf{r}}|i\rangle$ being the transition dipole moment of a two-level exciton subsystem, $e$ and $\hbar$ being the elementary charge and reduced Planck constant, and $\hat{\mathbf{r}}$ being the radius-vector operator. The quantity $\mathbf{E}_v(\mathbf{r}_0)$ is the *vacuum* electric field (the field per one photon) of the resonator mode at an exciton position. Thus, the coupling strength of the exciton and NC is described by the *coupling constant* ($g$), which defines two characteristic regimes of interaction and, hence, spontaneous emission. If the condition $g < (\gamma_{cav} + \gamma_{ex})/4$ is satisfied, the system is in the so-called *weak-coupling regime*. The expression in brackets is the total *decay rate* $\gamma = \gamma_{cav} + \gamma_{ex}$ ($\gamma_{ex}$ is decay rate of the exciton). In this regime, there is no coherent energy exchange between the molecule and nanocavity, and the variation of spontaneous emission can be described by the *Purcell factor* (see below). If the opposite condition is satisfied, $g > (\gamma_{cav} + \gamma_{ex})/4$, the system instead is *strongly coupled*. In this regime, the system is characterized by energy exchange between exciton and cavity (Rabi oscillations), resulting in a more complex behavior of spontaneous emission and splitting of the scattering spectrum (Rabi splitting) [65,152]. To describe the coherent scattering spectra of the system, the *coupled harmonic oscillator model* can be used [59,153]. This model considers a cavity driven by an external field, which is coupled to the exciton resonance with a coupling constant $g$.

To increase $g$ and achieve strong coupling, we can increase the dipole moment of the exciton subsystem and/or reduce the effective mode volume $V_{eff}$, which can be defined in the low loss limit as [154,155]

$$V_{eff} = \frac{\int \varepsilon(r)|E(r)|^2 \, dV}{\max\left(\varepsilon(r)|E(r)|^2\right)}, \qquad (2)$$

where $E(r)$ is the electric field of mode, $\varepsilon(r)$ is the permittivity distribution, and integration runs over the entire space including the NC and surrounding space.

As we discussed above, 1L-TMDCs have very strong dipole moment of excitons because of their direct bandgap and 2D nature. The effective mode volume is usually less than the physical volume of the resonant cavity and it strongly depends on its nature (plasmonic or dielectric). Plasmonic NCs can have effective mode volumes up to $\sim \lambda^3/10^4$ [53,144]. The mode volume of dielectric NCs are diffraction-limited: $V_{eff} \geq (\lambda/2n)^3$ and it can be shrunk by the increasing $n$. For c-Si nanoparticles with $n \approx 4$ in the visible range $V_{eff} \geq \lambda^3/512$.

Excitons in 1L-TMDCs have radiative $\gamma_{ex}^{rad}$ and absorptive $\gamma_{ex}^{loss}$ decay rates: $\gamma_{ex} = \gamma_{ex}^{rad} + \gamma_{ex}^{loss}$. In the weak-coupling regime, the decay rate $\gamma_{ex}$ can be enhanced via *Purcell effect*, which is quantitatively characterized by Purcell factor ($F_p$), expressed through the coupling constant $g$ or the quality factor Q [155,156]:

$$F_p \equiv \frac{\gamma_{ex}}{\gamma_{ex}^0} = \frac{g^2}{\gamma_{cav}\gamma_{ex}^0} = \frac{3}{4\pi^2}\left(\frac{\lambda}{n}\right)^3 \frac{Q}{V_{eff}}, \qquad (3)$$

with $\gamma_{ex}^0 = \frac{nd^2\omega^3}{3\pi\hbar\varepsilon_0 c^3}$ being the radiative decay rate in free space. Thus, the Purcell effect results in a modification of the spontaneous decay rate $\gamma_{ex}$ of an exciton induced by its interaction with the NC. This modification is significant if the NC is adjusted with the PL intensity maximum.

In general, the Purcell effect gives rise to enhancement of emission and dissipation rates, thus the Purcell factor can be divided into radiative ($F_p^{rad} = \gamma_{ex}^{rad}/\gamma_{ex}^0$) and dissipative ($F_p^{loss} = \gamma_{ex}^{loss}/\gamma_{ex}^0$). To characterize the fraction of energy emitted into photons, we also define the *quantum yield* (or *quantum efficiency*), defined as $\eta = \gamma_{ex}^{rad}/(\gamma_{ex}^{rad} + \gamma_{ex}^{loss})$. By introducing the vacuum quantum yield $\eta_0$ (see Section 2), the resulting emission quantum yield $\eta$ of an exciton coupled to a NC can be expressed through the Purcell factor: $\eta = \eta_0 \frac{F_p^{rad}}{\eta_0 F_p + (1-\eta_0)}$.

This equation says that the quantum yield can be enhanced by Purcell factor, which is important for 1L-TMDCs, because of their low radiation efficiency and decay times mismatch ($\gamma_{ex}^{rad} \ll \gamma_{ex}^{loss}$).

## 5. 1L-TMDCs-based nanostructures in weak and strong coupling regime

*5.1 Weak coupling regime*

Weak coupling regime in 1L-TMDCs based nanostructures manifests itself in increasing of the PL intensity without noticeable change in the spectrum. This increase relies on Purcell effect, which leads to quantum yield enhancement. The Purcell effect is caused by the mode volume (2) shrinking, and it depends on the NC Q-factor through its geometry and arrangement, as in Eq. (3). The so-called *gap plasmonic modes*, arising between nearby plasmonic particles, are of special interest because of their small mode volumes. In Ref. [54], monomer and dimer (supporting gap modes) Au NCs of four different sizes were examined for PL intensity enhancement from CVD-synthesized WS$_2$ flakes, Fig. 5(a). The figure shows that relatively small dimer NC formed by closely placed Au particles (75 nm) cause the largest enhancement of PL emission, whereas larger dimers have effects comparable to a single monomer.

Another exciting way to enhance the PL emission intensity is based on NCs designed to support two resonant (gap) modes at both 1L-TMDC PL emission and absorption maxima. In Ref. [53], a NC consisting of a silver (Ag) nanocube (~75 nm edge length) located over a gold film divided by a dielectric nanoscale spacer (<10 nm) has been studied [see the inset in Fig. 5(b)]. The dielectric spacer was filled by 1L-MoS$_2$. The NC supports two split modes at wavelengths 420 nm and 660 nm, with ultrasmall effective mode volumes of $\sim \lambda^3/10^3$. The modes resonant wavelengths have a fine overlap with the 1L-MoS$_2$ PL emission and absorption peaks. This spectral tuning and ultrasmall effective mode volumes of the NC resulted in 2000-fold enhancement in the PL intensity (normalized by unit area and emission of 1L-MoS$_2$ on SiO$_2$ substrate), Fig. 5(b).

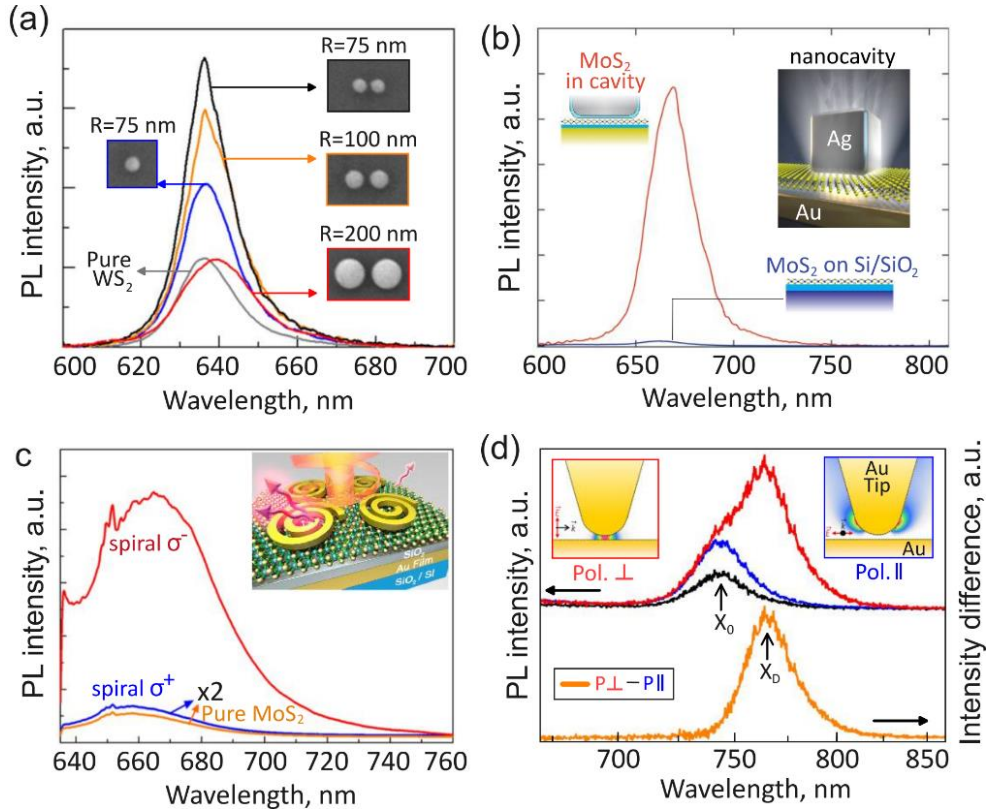

Fig. 5. 1L-TMDCs-based nanostructures in the weak coupling regime. PL intensity of CVD-grown 1L-WS$_2$ before and after fabrication of four different optical antennae. Insets are SEM images of each type of optical cavity; the scale bar is 400 nm. (b) PL spectra from a 1L-MoS$_2$ on a SiO$_2$/Si substrate (blue) and in the nanocavity (red) obtained using a diffraction-limited excitation spot. The intensity is measured per unit of excitation power and per unit of integration time. (c) PL spectra of the 1L-MoS$_2$ with and without spiral structures, under the excitation of different circular polarized light at 633 nm. Laser power was controlled at 2.1 mW. Inset: schematic of spiral ring structure on SiO$_2$/Au/SiO$_2$/Si substrate with circularly polarized light excitation. (d) Probing radiative emission of dark excitons of 1L-WSe$_2$ through polarization dependence of tip-enhanced photoluminescence. Excitation polarization dependent TEPL spectra of 1L-WSe$_2$ on an Au substrate at ~1 nm tip-sample distance with tip selective X$_D$ emission (orange curve). Inset: Finite-difference time domain (FDTD) simulation of the in-plane (right) and out-of-plane (left) optical field intensity and confinement. Adapted with permission from: (a) – Ref. [54], (b) – Ref. [53], (c) – Ref. [58], (d) – Ref. [24].

As we discussed in Section 2, 1L-TMDCs are of special interest because of their valley-selective circular dichroism, caused by lattice inversion symmetry breaking and strong spin-orbital coupling. Thus, the coupling of excitons belonging to different valleys in chiral NC, which interact differently with left ($\sigma^-$) and right ($\sigma^+$) circularly polarized light is of major interest for practical applications. An interesting geometry is a spiral ring resonant structure (Au) on SiO$_2$/Au/SiO$_2$/Si substrate, as presented in Fig. 5(c), inset. The 1L-MoS$_2$ is placed between spiral rings and the first SiO$_2$ spacer. The structure demonstrates strong dependence of PL intensity on the circular polarization handedness of the laser excitation. For example, for a left-handed 2-turn spiral rings the structure demonstrates 10-fold enhancement in PL intensity for $\sigma^-$ excitation at 633 nm wavelength, whereas the $\sigma^+$ excitation causes only a slight enhancement, Fig. 5(c).

In addition to conventional bright excitonic states, as discussed in Section 2, 1L-TMDCs possess spin-prohibited dark excitons with polarization in the out-of-plane direction, characterized by ultrahigh radiative lifetimes because of their low radiation loss. These states are particularly interesting for quantum information processing applications. It has been demonstrated that in 1L-TMDCs dark excitons can be excited through two-photon excitation spectroscopy [13] and in-plane magnetic field [72]. Additionally, emission from dark excitons can be detected if a 1L-TMDC is excited by a field with vanishing in-plane component, as it has been realized with SPP waves [157]. In Ref. [24], it was shown that ultrahigh Purcell effect can be used for dark exciton spectroscopy even at room temperature. Fig. 5(d) shows results from tip-enhanced PL spectroscopy of 1L-WSe$_2$ arranged on Au substrate. In-plane side excitation (field distribution is shown in the right inset) demonstrates enhancement in PL intensity (blue curve) over emission for high tip distances (black curve) with peak at the right exciton ($X_0$). More interestingly, out-plane side excitation (field distribution shown in the left inset) results in strong optical field confinement in the gap between the tip and the Au substrate. In this case, the PL peak is shifted to the dark exciton emission ($X_D$), with PL intensity enhancement $\sim 6\times10^5$-fold and a large Purcell factor $\geq 2\times10^3$.

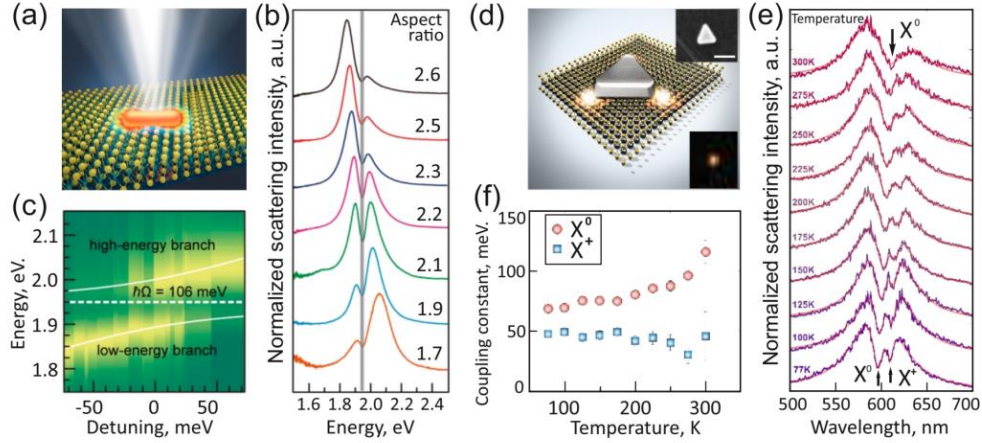

Fig. 6. 1L-TMDCs-based nanostructures in the strong coupling regime. (a) Schematic showing the heterostructure composed of an individual Au nanorod coupled to the WS$_2$. (b) Dark-field scattering spectra from different individual gold nanorods coupled to the same 1L-WS$_2$ flake (different curves correspond to various aspect ratios of the gold nanorods). (c) Colored coded normalized scattering spectra from the heterostructures with different detunings between the plasmon resonances and exciton. The two white lines represent the fittings using the coupled harmonic oscillator model. The dashed white line indicates the exciton transition energy. The Rabi splitting energy is determined as 106 meV. (d) Artist view of the hybrid system: high density of photonic states (hot-spots) is shown at the corners of the Ag nanoprism, which overlaps with the 1L-WS$_2$ for efficient plasmon-exciton interaction. Inset shows the SEM image of such particle (scale bar 100 nm) and a magnified view of the dark-field image. (e) Dark field spectra for an individual Ag nanoprism – 1L-WS$_2$ hybrid as a function of temperature. Semi-transparent red curves show coupled oscillator model fits of the data. (f) Temperature dependence of the coupling constants for exciton and trion extracted via coupled oscillator model. Adapted with permission from: (a) – Ref. [57], (b) – Ref. [17].

*5.2 Strong coupling regime*

Although structures supporting gap plasmons possess small effective mode volumes, they suffer from high dissipative losses, hindering the achievement of strong coupling regimes. Thus, for this purpose more carefully designed structures with low mode volumes and lower dissipative losses must be used [17,50,57,75,158]. Fig. 6(a) shows a possible layout, consisting of an individual gold nanorod placed on 1L-WS$_2$ (substrate is SiO$_2$/Si). Fig. 6(b) shows dark-field scattering spectra from different individual gold nanorods coupled to the same 1L-WS$_2$ flake for different nanorod aspect ratios, which allow to estimate the Rabi splitting energy for various frequency detunings. The resulting colored normalized scattering spectra from the hybrid structure with different detunings between plasmon resonances and the exciton is shown in Fig. 6(c). Splitting of high-energy and low-energy branches, with Rabi splitting energy of 106 meV, which satisfies the strong coupling condition, can be observed. Active control over strong coupling via temperature has also been observed.

As we discussed in Section 2, charged excitons (trions) can also be observed in 1L-TMDCs PL spectra at cryogenic temperatures or under 1L-TMDCs doping. The strong coupling with trions is of special interest because they form charged exciton-polaritons, which can be actively controlled by an external electrical bias. Charged exciton polaritons have been demonstrated in Ref. [17] at low temperatures. The structure is composed of Ag nanoprisms placed on 1L-WS$_2$ (substrate is SiO$_2$/Si). In this case, the electrical doping from the substrate was insignificant, hence only neutral ($X^0$) have been observed in PL spectra at room temperature, whereas the trion ($X^+$) has manifested itself at low temperatures, Fig. 6(e). The enhancement of trion coupling strength up to ~50 meV (the Rabi splitting is ~100 meV) at 77 K has been observed, Fig. 6(f). The observed reduction in exciton coupling strength with temperature increase can be explained by trion dissociation to excitons and free carriers.

## 6. Outlook and conclusion

We have reviewed state-of-the-art advances of hybrid exciton-polariton structures based on 1L-TMDCs coupled to plasmonic and dielectric nanocavities. First, we have discussed the optical properties of 1L-WS$_2$, WSe$_2$, MoS$_2$ and MoSe$_2$ materials, with special emphasis on their energy spectra, photoluminescence and absorption spectra, excitonic fine structure, and dynamics of exciton formation and valley depolarization. Then, we have provided a theoretical background to describe light-matter interaction in such hybrid exciton-polariton structures, and we have applied it to discuss weak and strong coupling regimes in 1L-TMDCs-based exciton-polariton systems.

This research area is still very young, yet full of exciting promises and opportunities. Several more studies and experiments are needed to fully understand the nature of this new material platform, and its potential for new technology. We envision several future research directions in this context. First, although all-dielectric NCs [131] possess attractive properties, like low dissipative losses and strong optical magnetic response (Section 3), there are only preliminary results in the area of their interaction with 1L-TMDCs [75]. Interaction of 1L-TMDCs with anopole-like [150] field configurations and non-radiating resonant eigenstates and bound states in the continuum [159,160] can offer exciting opportunities. Second, various applications require light emitting systems containing 1L-TMDCs to have highly directive (and even steerable) spontaneous emission patterns, for instance, for the emission collection enhancement. This issue is addressed in only a few recent papers [61,161] and it requires additional studies. In this context, suitably designed resonant nanostructures or metasurfaces can spatially separate valley polarized excitons, which may bring valleytronics to real-life applications [162].

## Acknowledgments

The authors are thankful to Mr. Mingsong Wang, Dr. Oleg Kotov, and Dr. Denis Baranov for discussions. This material is based upon work supported by the Air Force Office of Scientific Research under award number FA9550-17-1-0002.